\documentclass[twocolumn,showpacs,preprintnumbers,amsmath,amssymb]{revtex4}

\usepackage{mathptmx} 
\usepackage{times} 
\usepackage{amsmath} 
\usepackage{amssymb}  

\newcommand{\p}{\partial}

\newcommand{\sd}{Schr\"{o}dinger }
\newcommand{\pt}{P\"{o}schl-Teller }

\newcommand{\A}{\mathcal{A}}
\newcommand{\B}{\mathcal{B}}
\newcommand{\C}{\mathcal{C}}
\newcommand{\G}{\mathcal{G}}
\newcommand{\h}{\mathcal{H}}

\newcommand{\la}{\mathcal{L}}
\newcommand{\Dw}{{\mathcal{D}_\omega}}
\newcommand{\Ds}{{\mathcal{D}_\infty}}

\newcommand{\M}{\mathbf{M}}
\newcommand{\R}{\mathcal{R}}

\newtheorem{theorem}{Theorem}

\newtheorem{definition}{Definition}
\newtheorem{remark}{Remark}

\begin{document}
\title{Controllability of Scattering States of Quantum Mechanical Systems}

\author{Re-Bing Wu}
 \altaffiliation{Department of Chemistry, Princeton
University, Princeton, NJ 08544, USA} \email{rewu@Princeton.EDU}

\author{Tzyh-Jong Tarn}%
\altaffiliation{Department of Electrical and Systems Engineering,
Washington University, St. Louis, MO 63130, USA}
\email{tarn@wuauto.wustl.edu}

\author{Chun-Wen Li}
\altaffiliation{Department of Automation, Tsinghua University,
Beijing, 100084, P. R. China} \email{lcw@mail.tsinghua.edu.cn}

\date{\today}
\begin{abstract}
This paper provides a framework for the control of quantum
mechanical systems with scattering states, i.e., systems with
continuous spectra. We present the concept and prove a criterion
of the approximate strong smooth controllability. Our results make
non-trivial extensions from quantum systems with finite
dimensional control Lie algebras to those with infinite
dimensions. It also opens up many interesting problems for future
studies.
\end{abstract}

\maketitle

Recently, increasing research has been done on quantum mechanical
systems with scattering states that possess practical applications
\cite{lloyd4,contqc,alhassid6,rice3}. This has been a
long-standing problem in the area of molecular control, where the
attempts to breaking chemical bonds naturally evolve into various
topics on discrete-continuum transitions of molecular states
\cite{alhassid6,rice3,shapiro1}, i.e. photon-dissociation(from
discrete to continuum), photon-association(from continuum to
discrete) and laser catalysis (from continuum to continuum).
Another prominent motivation is the continuous quantum computer
\cite{lloyd4,contqc} that processes quantum information over
continuous spectra. Serious theoretical studies have proved that
they might be more sufficient in some tasks compared with their
discrete counterparts.

These problems naturally fall under the system control theory
\cite{tarn2,tarn5,lan2,rabitz3,rabitz1,rabitz2,rabitz6}. To the
authors' knowledge, most existing studies focus on finite
dimensional bound-state quantum systems that have been strikingly
successful in practice \cite{rabitz3,dominico1,schirmer}. Very
little has been done to scattering-state quantum control systems
\cite{tarn5,lan2,slemrod,brockett1} except some specialized
discussions in the molecular control \cite{shapiro1}, where the
calculation of controlled discrete-continuum transition
probabilities are carried out along the lines of perturbation
theory and adiabatic approximations under weak field assumptions.

Amongst various topics on quantum control, the controllability is
of fundamental importance in understanding the physical mechanisms
of quantum control \cite{rabitz3,tarn2,rabitz6}. In this paper, we
are investigating the controllability for a scattering-state
quantum system that possesses an infinite dimensional control Lie
algebra. As is well known in mathematics, the extension from
finite dimensional Lie algebras to infinite dimensional Lie
algebras is not trivial. Theoretically, the infinite
dimensionality is the key to break through the {\it HTC No-Go}
Theorem \cite{tarn2} so as to enable the ambitious control over a
set of all scattering states. In these regards, our works make a
quantum leap over the earlier result on finite dimensional
systems, and that of \cite{tarn2,lan2}, which, although embraces
the cases of both discrete\cite{tarn2} and continuous
spectra\cite{tarn5}, is still limited by the finite dimensionality
of control Lie algebras.

Consider quantum control system in the form of the following \sd
equation:

\begin{equation}\label{abc'}
  i\hbar{\p\over\p{t}}\psi(t)=\left[H_0'+\sum_{j=1}^m{u_j(t)H_j'}\right]\psi(t),\,\psi(0)=\psi_0.
\end{equation}

\noindent where the controls $u_j(t)$ are real piecewise constant
functions of time. The quantum state $\psi(t)$ evolves in a
separable Hilbert space $\h$ that carries a unitary representation
space of the symmetry algebra. The Hamiltonians
$H_0',H_1',\cdots,H_m'$ are Hermitian operators acting on $\h$.
For convenience, we rewrite (\ref{abc'}) with skew-Hermitian
operators $H_i=H_i'/(i\hbar)$, i.e.,
\begin{equation}\label{abc}
 {\p\over\p{t}}\psi(t)=\left[H_0+\sum_{j=1}^m{u_j(t)H_j}\right]\psi(t),\,\psi(0)=\psi_0.
\end{equation}

In order to facilitate the analysis of system structures, we
choose to embed the system in an algebraic framework based on an
associated intrinsic symmetry algebra, say
$\la=\{L_1,\cdots,L_d\}_{LA}$, of the quantum system under
consideration. The subscript "{\it LA}" denotes the Lie algebra
generated by the finite number of operators in the curly bracket.
We assume that the Hamiltonians can be formally expressed as
elements in the universal enveloping algebra $E(\la)$
\cite{barut}, i.e. in terms of polynomials of the generators of
$\la$. Obviously, the controllability Lie algebra
$\A=\{H_0,H_1,\cdots,H_m\}_{LA}$ is a Lie subalgebra of $E(\la)$.

The algebraic modelling method covers a wide class of quantum
control systems such as harmonic oscillator \cite{tarn2} and
hydrogen atom \cite{lan2}. A paradigm closely related to this
paper is the continuous quantum computer modelled as
\cite{lloyd4}:
\begin{equation}\label{lloyd}
  \begin{array}{rl}
    i\hbar{\p\over\p{t}}\psi(x,t) =&[(p^2+x^2)+u_1p+u_2x+u_3(x^2+p^2) \\
     & +u_4(xp+px)+u_5(x^2+p^2)^2]\psi(t), \\
  \end{array}
\end{equation}
\noindent where $p=-i\hbar\p_x$, $[x,p]=i\hbar$. Here the
Heisenberg algebra $h(1)=\{x,p,i\}_{LA}$ plays the role of the
symmetry algebra of the system. The Hamiltonians for quantum
computation are derived from quantization of classical variables
of a harmonic oscillator into the universal enveloping algebra
$E(h(1))$ \cite{tarn1,aldaya}. Comparing this system with the
example of harmonic oscillator in \cite{tarn2}:
\begin{equation}\label{lloyd'}
  \begin{array}{rl}
    i\hbar{\p\over\p{t}}\psi(x,t) =&[(p^2+x^2)+u_1p+u_2x], \\
  \end{array}
\end{equation}
one can observe the distinction between these two systems: the
controllability algebra of (\ref{lloyd}) is infinite dimensional,
while that of (\ref{lloyd'}) is finite dimensional. As indicated
in \cite{lloyd4}, the infinite dimensionality is necessary for
universality of quantum computation using this model.

The symmetry algebras of scattering-state quantum mechanical
systems are necessarily noncompact. From the theory of unitary
representation, $\h$ has to be an infinite dimensional Hilbert
space in which the quantum state prevails \cite{barut,peres}. Also
due to the noncompactness, the unboundedness of Hamiltonians
brings severe domain constraints \cite{barut}. Thus one has to
specify a dense subset of $\h$ on which the Hamiltonians are
well-defined, invariant and the state-evolution can be expressed
globally in exponential form. For systems with finite dimensional
control Lie algebras , Huang, Tarn and Clark \cite{tarn2}
suggested the analytic domain(the existence is ensured by Nelson's
theorem \cite{nelson})
$$\Dw=\left\{\omega\in\h:\,\sum_{n=0}^\infty{\frac{1}{n\,!}\sum_
{1\leq{i_1,\cdots,i_n}\leq{d}}{\|H_{i_1}\cdots{H_{i_n}}\omega\|s^n}}<\infty\right\}$$
\noindent as a candidate, on which they present the notion of
analytic controllability. The state-evolution of these systems are
therefore restricted on a locally finite dimensional manifold with
finite directions to move towards.

However, there does not exist, in general, a dense analytic domain
for $\A$ as an infinite dimensional Lie subalgebra of $E(\la)$.
Nevertheless, as implied in \cite{omori}, the existence of
analytic domain is not necessary for the exponential formula. A
natural choice is to enlarge the analytic domain to the collection
of differentiable vectors of the associated symmetry algebra
$\la$:

$$\Ds=\{\phi\in\h : \|L_1^{s_1}\cdots
L_d^{s_n}\phi\|<\infty;
s_i=0,1,2,\cdots\}.\\
$$

Given an insight into the space structure, $\Ds$ forms a complete
Frechet space endowed with a locally convex topology that makes
the Hamiltonians continuous\cite{omori,madrid}. With this
topology, the control algebra $\A$ can be exponentiated to form a
Frechet (or ILH-) Lie transformation group $\G$
\cite{omori,onishchik} acting on $\Ds$. Parallel with the analytic
domain, we call $\Ds$ the smooth domain. For instance, the smooth
domain for (\ref{lloyd}) is the Schwartz space

$$\left\{v(x)\in{L^2(\mathbb{R})}:\sup_{\alpha,\beta\geq{0}}\left|\,x^\alpha\left(\frac{d}{dx}\right)^\beta{v(x)}\right|<\infty\right\}.$$

\begin{remark}
The smooth domain provides a natuaral characterization of
scattering states. Denote the dual space of $\Ds$ by $\Ds^*$, we
obtain a triad of linear spaces, i.e., the rigged Hilbert space
\cite{madrid}:
$$\Ds\subset\h\subset\Ds^*,$$
\noindent with $\Ds$ dense in $\h$ and $\h$ dense in $\Ds^*$.
While $\Ds$ contains all experimentally preparable states for
(\ref{abc}), the scattering states can be specified as vectors in
$\Ds^*$ which have run out of the Hilbert space $\h$. Since $\Ds$
is dense in $\Ds^*$, the system can approach arbitrarily close to
any scattering state if all states in the smooth domain is
reachable. From this viewpoint, we can make sense of the dynamical
control over {\it nonphysical} scattering states.
\end{remark}

The smooth domain and the exponentiated Frechet Lie group have
already laid a technical basis in system analysis of the
controllability properties. Let $\M$ be the closure of the set

$$\begin{array}{r}
\{\exp(s_1H_{\alpha_1})\cdots\exp(s_kH_{\alpha_k})\psi_0| s_k\in\mathbb{R},  \\
\alpha_k=0,1,\cdots,m,k\in\mathbb{N}\}   \\
\end{array}$$

\noindent for the initial state $\psi_0\in\Ds$. Note $\M$ is
allowed to be an infinite dimensional sub-manifold of $S_\h$, the
unit sphere in $\h$. Apparently, $\M$ is the maximal set of
possibly reachable states from $\psi_0$. Here we give the
definition of controllability to be studied in this paper:
\begin{definition}[Smooth Controllability]
Denote $\R(\psi)$ as the reachable set of the initial quantum
state $\psi_0\in\M\cap\Ds$ by all quantum states that can be
driven from $\psi_0$ under properly adjusted controls. Quantum
mechanical control system (\ref{abc}) is said to be approximately
smoothly controllable if $\overline{\R(\psi)}=\M\cap\Ds$ (the
closure is with respect to the locally convex topology of $\Ds$).
Moreover, the system is said to be approximately strongly smoothly
controllable if the closure of reachable set $\R_t(\psi)$ at any
time $t>0$ equals to $\M\cap\Ds$.
\end{definition}

\begin{remark}
The prevailing definition of quantum controllability implicitly
assigns the manifold $\M$ to $S_\h$, the unit sphere in $\h$. This
definition can be taken as a special case of that given here. It
is another interesting and important problem to investigate when
the reachable set of a controllable system covers $S_\h$.
\end{remark}

Now we are standing at the same starting point as in previous
studies, where the existing results \cite{tarn2} can be naturally
extended to quantum mechanical systems with infinite dimensional
control algebras. Note that only finite control parameters are
available to affect the state-evolution, repeated switching
operations can generate only finite directions within any finite
time interval. Therefore, at most {\it approximate}
controllability can be achieved over an infinite dimensional
manifold $\M$ under piecewise constant controls. Nevertheless,
from a practical point of view, approximate controllability is
already enough for most situations. The following is our main
result:

\begin{theorem}\label{ac}
Let the Lie algebra $\B=\{H_1,\cdots,H_m\}_{LA}$ and
$\C=\{ad_{H_0}^jH_i,\,j=0,1,\cdots$, $i=1,\cdots,m\}_{LA}$, where
$ad_{H_0}^0H_i=H_i,\,ad_{H_0}^{j}H_i=[H_0,ad_{H_0}^{j-1}H_i]$. The
system (\ref{abc}) is approximately strongly smoothly controllable
if the following conditions are satisfied:
\begin{enumerate}
    \item $[\,\B,\C\,]\subseteq{\B}$;
    \item For any $\phi\in\M\cap\Ds$, $\C(\phi)=\A(\phi)$ and they
    are infinite dimensional.
\end{enumerate}
\end{theorem}

{\noindent\bf proof} Due to the space limitation, we present a
condensed proof in this paper. Recall that the primary obstacle
for controllability analysis is the presence of free evolution
which can not been artificially adjusted. From a system
theoretical point of view, the basic idea is to find out enough
"adjustable" flows generated by interactions between the system
Hamiltonians, so that the free evolution governed by $H_0$ can be
cancelled and recreated. This in turn leads to the approximate
strong smooth controllability prevailing over $\M\cap\Ds$.

We need to implement this idea in a strict way. A Hamiltonian $X$
is said to be strongly attainable if its integral curve passing
any $\psi_0\in\M\cap\Ds$ is contained in the closure of the
infinitesimal-time reachable set
$\R_0(\psi_0)=\bigcap_{t>0}\R_{\leq t}(\psi_0)$, where $\R_{\leq
t}(\psi_0)$ denotes the reachable set within $t$ units of time.
The collection of strongly attainable Hamiltonians form a Lie
algebra, i.e. they are closed under linear combination and
commutation operations. This novel property is obvious from the
well-know Trotter's formula \cite{davis}:

$$\begin{array}{rcl}
  \exp s(X+Y)\phi & = & \lim_{n\rightarrow\infty}[\exp (sX/n) \exp
(sY/n)^n]\phi, \\
  \exp s[X,Y]\phi & = & \lim_{n\rightarrow\infty}[\exp
(\sqrt{s/n}X) \exp
 (\sqrt{s/n}Y)  \\
   & & \exp(-\sqrt{s/n}X)\exp(-\sqrt{s/n}Y)]^n\phi. \\
\end{array}$$

Now we seek strongly attainable Hamiltonians. Apparently, the
control Hamiltonians $H_1,\cdots,H_m$ are strongly attainable,
because they dominate the system evolution if we tune the controls
sufficiently large (see detailed proof in \cite{wrb}). Thus the
spanned Lie algebra $\B$ is strongly attainable.

Subsequently, we investigate the interaction of a strongly
attainable Hamiltonian $H\in\B$ with the free Hamiltonian $H_0$.
From the Campbell-Baker-Hausdorff formula \cite{onishchik}

$$\begin{array}{cl}
   & \exp (tH) \exp sH_0 \exp
(-tH)\phi \\
  = & \exp\left\{s(H_0+t[H,H_0])+\frac{t^2}{2!}[H,[H,sH_0]]+\cdots\right\}\phi, \\
\end{array}$$

\noindent where $\phi\in\M\cap\Ds$, we observe that the higher
order terms on the right hand side are all strongly attainable
Hamiltonians in $\B$ under the condition
$[\,\B,\C\,]\subseteq{\B}$. So we have

$$\exp s(H_0+t[H,H_0])\psi_0\subseteq \overline{\R_s(\psi_0)},\,s>0,$$

\noindent by which we derive that
$$\begin{array}{rcl}
\exp (\pm s[H,H_0])\psi_0 &=&  \lim_{t\rightarrow\pm\infty}\exp\frac{s}{|t|}(H_0+t[H,H_0])\psi_0\\
   &\subseteq& \lim_{t\rightarrow\pm\infty}\overline{\R_{\frac{s}{|t|}}(\psi_0)}=\overline{\R_0(\psi_0)}. \\
\end{array}
$$

Thus we proved that the first-order term $[H,H_0]$ is strongly
attainable. Inductively by repeated commutation operations, the
system evolution can be guided along any flows generated by
Hamiltonians in $ad_{H_0}^k\B$ for every integer $k$ with
arbitrary precision. Therefore, the Lie algebra $\C$ is strongly
attainable. This is to say, the integral manifold $(\exp \C)
\psi_0$ of $\C$, which coincides with $\M$ under the condition
$\C(\phi)=\A(\phi)$ for any $\phi\in\M\cap\Ds$, is contained in
$\overline{\R_0(\psi_0)}$.

Therefore, the system can be steered arbitrary closely to any
state in $\M\cap\Ds$ at any time $t>0$, because

\begin{equation}
\begin{array}{ll}
  \M\cap\Ds & \subseteq\exp(tH_0)(\exp \A)\psi_0=\exp(tH_0)(\exp
\C)\psi_0 \\
   & \subseteq\overline{\R_t(\psi_0)}\subseteq\M\cap\Ds. \\
\end{array}
\end{equation}

Thus $\overline{\R_t(\psi_0)}=\M\cap\Ds$, i.e., the system is
approximately strongly smoothly controllable over $\M$.

Physically, our result enables one to use finite controls to
modulate the superposition of {\it innumerable} scattering states
so as to resist the irreversible dispersion of wavepacket of
scattering-state quantum systems. The fact is somewhat amazing and
has been earlier questioned by Zhao and Rice \cite{rice3} for the
presence of chaos. Actually, as they also recognized later, the
essence of controllability is not altered because controllability
problems always concern themselves with evolution on finite time
internals, although the presence of chaos do make the control much
more complicated.

To illuminate the ideas presented in this paper, we will discuss
an example described in \cite{tarn5}. The example is typically
associated with an infinite dimensional control algebra. The
algebraic model characterizes the scattering states of particles
subject to \pt potentials based on a scattering algebra
$so(2,1)=\{L_x',L_y',L_z'\}_{LA}$ with $L_y'$ compact and
$L_x',L_z'$ noncompact (see \cite{tarn5} for detail), which
commutation relations read:

$$[L'_x,L'_y]=-iL'_z,\,[L'_y,L'_z]=-iL'_x,\,[L'_z,L'_x]=iL'_y.$$

We consider a control system as follows:

\begin{equation}\label{pt}
i\hbar{\p\over\p{t}}\psi(t)=[a{L'}_z^2+u_1L'_x+u_2L'_y+u_3{L'}_x^2]\psi(t).
\end{equation}

The noncompact free Hamiltonian $a{L'}_z^2$ generates a continuous
spectrum. Let $|j,m\rangle$ be the simultaneous eigenvectors of
the compact generator ${L'}_y$ and the Casimir operator
$C={L'}_y^2-{L'}_x^2-{L'}_z^2$, the smooth domain of (\ref{pt}) is

\begin{equation}
\Ds=
\{x=\sum_{m=j}^\infty{x_m|j,m\rangle}\,|\,\lim_{|m|\rightarrow\infty}m^nx_m=0,\,\forall\,n\in\mathbb{N}\}
\end{equation}

\noindent as described in \cite{lindblad}.

It is not difficult to verify inductively that any element in
$E(so(2,1))$, including the free Hamiltonian, may be generated by
repeated commutations and linear combinations of the control
Hamiltonians. Hence the Lie algebras $\A,\B,\C$ coincide with
$E(so(2,1))$. Let $\M$ be the integral manifold of $\A$ through
$\psi_0$. According to Theorem \ref{ac}, strong approximate smooth
controllability follows on $\M$. We conjecture that $\M$ is dense
in the unit sphere $S_\h$, but the rigorous proof has not been
found.

From the above derivation, the first-order control Hamiltonians
manifest fundamental significance for shaping the quantum
wavepacket by modulating the interaction of control Hamiltonians.
At the first glance, the operators $K_\pm=L_x\pm{L_y}$ resemble
the ladder operators of harmonic oscillators which is well-known
for "raising" and "lowering" of energy levels. But intuitively,
the "ladder" operators are not allowed to generate any discrete
shift of levels on a continuous spectrum. One can find by simple
calculations that $K_\pm$ shift eigenvalues of $L_z$ by $\pm i$
units \cite{mukunda}, which is of course absurd because the
Hermitian $L_z$ has a real continuous spectrum. Indeed, the
contradiction arises from the fact that $K_\pm$ can only operate
on the superposition of scattering states, but illegally upon the
scattering states that are outside the Hilbert space
\cite{mukunda}. Interested readers are referred to \cite{mukunda}
for more details.

The use of high-order operators in the universal enveloping
algebra enables one to expand an infinite dimensional algebra by
finite control Hamiltonians, so that the quantum state can be
driven to an infinite dimensional manifold. Thus the high-order
terms are important to enhance the control of quantum systems. As
shown in the optical scheme for continuous quantum computation
\cite{lloyd4}, the key operation is chosen as the nonlinear Kerr
process $(x^2+p^2)^2$ in equation (\ref{lloyd}) as a higher order
terms. This high order term plays an essential role in achieving
universality of quantum computation over continuous variables.

In conclusion, this paper provides a clearer understanding of
quantum control over infinite dimensional manifolds of scattering
states. Our result advances the existing results to a broader
landscape. Back to the cases of finite dimensional control
algebras, the extension of analytic controllability to the larger
smooth domain, which has been conjectured in \cite{tarn2}, can be
taken as a trivial corollary of Theorem \ref{ac}. The extension to
infinite dimensional manifolds makes it possible to explore full
control over the whole set of scattering states. It opens up many
interesting problems for study, e.g. the control between discrete
and continuous spectra and the control of resonances.

\section{ACKNOWLEDGMENTS}
This research was supported in part by the National Natural
Science Foundation of China under Grant Number 60433050 and
60274025. TJT would also like to acknowledge partial support from
the U.S. Army Research Office under Grant W911NF-04-1-0386.


\begin{thebibliography}{99}

\bibitem{lloyd4}
S.~Lloyd and S.L. Braunstein.
\newblock Quantum computation over continuous variables.
\newblock {\em Phy. Rev. Lett.}, 82:1784--1787, 1999.

\bibitem{contqc}
S.~L. Braunstein and A.~K. Pati, editors.
\newblock {\em Quantum Information with Continuous Variables}. Kluwer
  Academics, 2003.

\bibitem{alhassid6}
Y.~Alhassid, J.~Engel, and F.~Iachello.
\newblock Algebraic approach to dissociation from bound states.
\newblock {\em Phy.Rev.Lett.}, 57:9--12, 1986.

\bibitem{rice3}
M.~Zhao and S.A. Rice.
\newblock Comment concerning the optimum control of transformations in an
  unbounded quantum system.
\newblock {\em J.Chem.Phys.}, 4:2465--2472, 1991.

\bibitem{shapiro1}
M.~Shapiro and P.~Brumer.
\newblock {\em Principles of the quantum control of molecular processes}.
\newblock Wiley-Interscience, U.S., 2003.

\bibitem{tarn2}
G.~M. Huang, T.~J. Tarn, and J.~W. Clark.
\newblock On the controllability of quantum-mechanical systems.
\newblock {\em J. Math. Phys.}, 24:2608--2618, 1983.

\bibitem{tarn5}
T.~J. Tarn, J.~W. Clark, and D.~G. Lucarelli.
\newblock Controllability of quantum mechanical systems with continuous
  spectra.
\newblock In {\em Proceedings of the 39th IEEE Conference on Decision and
  Control}, pages 943--948, Sydney, 2000.

\bibitem{lan2}
C.H. Lan, T.J. Tarn, Q.S. Chi, and J.W. Clark.
\newblock Strong analytic controllability for hydrogen control systems.
\newblock In {\em Proceedings of the 43rd IEEE Conference on Decision and
  Control}, 2004.

\bibitem{rabitz3}
V.~Ramakrishna, M.~V. Salapaka, M.~Dahleh, H.Rabitz, and
A.~Peirce.
\newblock Controllability of molecular systems.
\newblock {\em Phy. Rev. A}, 51:960--966, 1995.

\bibitem{rabitz1}
W.~S. Warren, H.~Rabitz, and M.~Dahleh.
\newblock Coherent control of quantum dynamics: the dream is alive.
\newblock {\em Science}, 259:1581--1588, 1993.

\bibitem{rabitz2}
H.~Rabitz, R.~de~Vivie-Riedle, M.~Motzkus, and K.~Kompa.
\newblock Wither the future of controlling quantum pheonomena?
\newblock {\em Science}, 288:824--828, 2000.

\bibitem{rabitz6}
H.~Rabitz, M.~Hsieh, and C.~Rosenthal.
\newblock Quantum optimally controlled transition landscapes.
\newblock {\em Science 303 (2004):}, 303:1998--2001, 2004.

\bibitem{dominico1}
D.D'Alessandro.
\newblock Topological properties of reachable sets and the control of quantum
  bits.
\newblock {\em System and Control Letters}, 41:213--221, 2000.

\bibitem{schirmer}
S.G. Schirmer, I.C. Pullen, and A.I. Solomon.
\newblock Identification of dynamical ie algebras for finite-level quantum
  control systems.
\newblock {\em J. Phys. A}, 35:2327--2340, 2002.

\bibitem{slemrod}
J.~E. Slemrod.
\newblock Controllability for a class of nondiagonal hyberbolic distributed
  bilinear systems.
\newblock {\em Appl. Math. Optim.}, 11:57--76, 1984.

\bibitem{brockett1}
R.W. Brockett, C.~Rangan, and A.M. Bloch.
\newblock The controllability of infinite quantum systems.
\newblock In {\em Proceedings of the 42th IEEE Conference on Decision and
  Control}, pages 428--433, Hawaii, U.S.A., 2003.

\bibitem{barut}
A.~O. Barut and R.~Raczka.
\newblock {\em Theory of group representations and applications}.
\newblock Polish Scientific Publishers, Wasazawa, 1980.

\bibitem{tarn1}
T.~J. Tarn, G.~M. Huang, and J.~W. Clark.
\newblock Modelling of quantum mechanical control systems.
\newblock {\em Mathematical Modelling}, 1:109--121, 1980.

\bibitem{aldaya}
V.~Aldaya and J.A. Azcarraga.
\newblock Quantization as a consequence of the symmetry group: An approach to
  geometric quantization.
\newblock {\em J. Math. Phys.}, 23:1297--1305, 1999.

\bibitem{peres}
A.~Peres.
\newblock {\em Quantum Theory: Concepts and Methods}.
\newblock Kluwer Academic Publishers, New York, 1998.

\bibitem{nelson}
E.~Nelson.
\newblock Analytic vectors.
\newblock {\em Annals of Mathematics}, 70:572--615, 1959.

\bibitem{omori}
H.~Omori.
\newblock {\em Infinite dimensional {L}ie groups}.
\newblock American Mathematical Society, U.S.A., 1997.

\bibitem{madrid}
R.~Madrid.
\newblock Rigged {H}ilbert space treatment of continuous spectrum.
\newblock {\em Fortschr. Phys.}, 2:185--216, 2002.

\bibitem{onishchik}
A.L. Onishchik(Ed.).
\newblock {\em Lie groups and Lie algebras {\rm I}}.
\newblock Springer-Verlag, Berlin, 1993.

\bibitem{davis}
E.~B. Davis.
\newblock {\em One-parameter semigroups}.
\newblock Academic Press, London, 1980.

\bibitem{wrb}
R.~B. Wu.
\newblock {\em Symmetry based modelling and control of infinite dimensional
  quantun mechanical control systems}.
\newblock PhD thesis, Tsinghua University, P.R. China, 2003.
\newblock in Chinese.

\bibitem{lindblad}
G.~Lindblad and B.~Nagel.
\newblock Continuous bases for unitary irreducible representations of
  $su(1,1)$.
\newblock {\em Ann. Inst. Henri Poincar$\acute{e}$}, XIII:27--56, 1970.

\bibitem{mukunda}
N.~Mukunda.
\newblock Unitary representation of the group $O(2,1)$ in an $O(1,1)$ basis.
\newblock {\em J. Math. Phys}, 8:2210--2220, 1967.

\end{thebibliography}
\end{document}